\documentclass[12pt,a4paper,final,floatfix]{iopart}

\usepackage{iopams}
\usepackage{graphicx}
\expandafter\let\csname equation*\endcsname\relax
\expandafter\let\csname endequation*\endcsname\relax
\usepackage{amsmath}
\usepackage[breaklinks=true,colorlinks=true,linkcolor=blue,urlcolor=blue,citecolor=blue]{hyperref}

\newcommand{\be}{\begin{eqnarray}}
\newcommand{\ee}{\end{eqnarray}}
\newcommand{\bfa}{{\bf a}}

\newcommand{\bfr}{{\bf r}}

\newcommand{\bfp}{{\bf p}}
\newcommand{\bfk}{{\bf k}}
\newcommand{\bfu}{{\bf u}}

\newcommand{\bfl}{{\bf l}}

\newcommand{\bfn}{{\bf n}}
\newcommand{\bfm}{{\bf m}}

\newcommand{\wbe}{\begin{widetext}}
\newcommand{\wee}{\end{widetext}}

\begin{document}

\title{Many-Body Formation and Dissociation of a Dipolar Chain Crystal}

\author{Jhih-Shih You$^{1,2,3}$ and Daw-Wei Wang$^{1,2}$}
\address{$^1$Physics Department and Frontier Research Center on Fundamental and Applied Sciences of Matter, National Tsing-Hua University, Hsinchu,
Taiwan}
\address{$^2$Physics Division, National Center for Theoretical Sciences, Hsinchu, Taiwan}
\address{$^3$Department of Physics, University of California, San Diego, CA 92093}
\ead{jhihshihyou@gmail.com}

\begin{abstract}
We propose an experimental scheme to effectively assemble chains of dipolar gases with an uniform length in a multi-layer system. The obtained dipolar chains can form a chain crystal with the system temperature easily controlled by the initial lattice potential and the external field strength during process. When the density of chains increases, we further observe a second order quantum phase transition for the chain crystal to be dissociated toward layers of 2D crystal, where the quantum fluctuation dominates the classical energy and the compressibility diverges at the phase boundary. Experimental implication of such dipolar chain crystal and its quantum phase transition is also discussed.
\end{abstract}

\vspace{2pc}
%\submitto{\NJP}
%% Comment out if separate title page not required
%\maketitle

\section{Introduction}

Recently tremendous interest in exotic strongly-correlated physics
arises from the anisotropic long-ranged dipolar interactions, which can be realized in ultracold polar molecules \cite{Ni,Deiglmayr0,Aikawa}, magnetic atoms of large electronic spin \cite{Griesmaier,Beaufils,Ferlaino,Lev}
and highly excited Rydberg atoms\cite{Walker,Bloch1}. These systems provide new avenue for ultracold physics beyond those with short-range and isotropic interactions of ordinary atomic gases, offering novel quantum phases that cannot be achieved even in traditional
condensed matter systems\cite{Menotti,Baranov}. One of the most interesting systems along this line is the self-assembled dipolar chain liquid in an array of one- or two-dimensional structure \cite{DW,Zhu,Pupillo,Klawunn,Potter,Armstrong0,Volosniev,Armstrong}, where a strong optical lattice potential suppresses the possible collapse instability in the attractive interaction side\cite{Ospelkaus, Santos}. The stabilized composite particles made of dipolar atoms/molecules in such multi-component systems hybridize the few-body and many-body physics and can be a quantum counter-part of the colloid fluid, which has been extensively studied in soft-matter physics\cite{deGennes, Teixeira, Butter, Yethiraj, Klokkenburg}.

%----------------
\begin{figure}[htbp!]
\centering
\includegraphics[width=9cm]{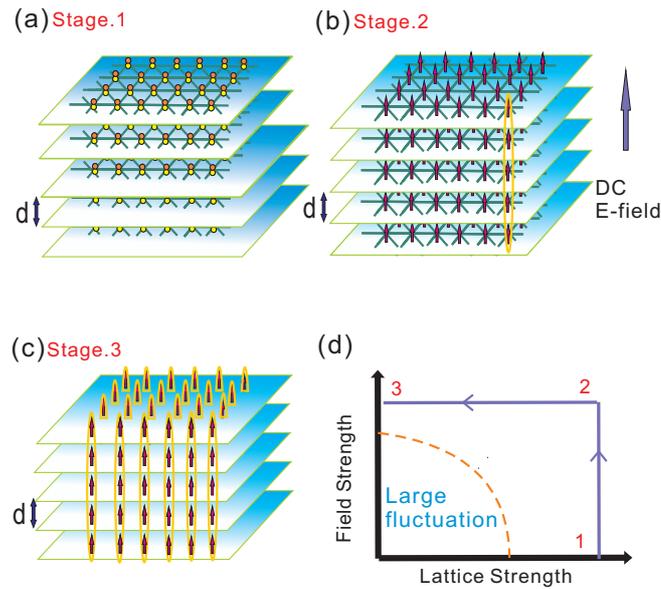}
\caption{ Three stages to form a dipolar chain crystal: (a) Prepare polar molecules in a Mott Insulator state in a 3D optical lattice. (b) Polarize the dipole moment by an external electric or magnetic field to form chains.(c)
Adiabatically switch off the in-plane optical lattice to form a chain liquid or a chain crystal. (d) Three stages in a terms of external parameters (lattice strength v.s. external field). Dashed Circle indicates the regime of large spatial fluctuation, which is avoided completely in this work.
}\label{fig1}
\end{figure}
%----------------
It is known that energy of a single dipolar chain is lower than that of unbound free dipoles even in small dipole moment limit \cite{DW}. However, the collision rate for such many particles to form such a bound state can be still low in the current experimental parameter regime. How to develop an experimentally realistic scheme to increase the collision rate is therefore an important issue for realizing a dipolar chain liquid. Besides, most studies in the literature did not include the interaction between chains, so the general many-body properties of such interesting systems are still unclear \cite{DW,Zhu,Pupillo,Klawunn,Potter,Armstrong0,Volosniev,Armstrong}.

In this paper we first propose a scheme to experimentally generate a dipolar chain liquid by manipulating a 3D optical lattices and the DC external field (see figures \ref{fig1}(a)-(d)), and investigate its many-body properties in the regime of crystal phase (i.e. strongly interacting regime). In order to have a chain liquid (or crystal) of almost the same length, one can first prepare dipolar gases (for example, polar molecules \cite{Danzl,Chotia,Hazzard}) in a Mott insulator state with one particle per site in a 3D optical lattice. An external DC electric field is then adiabatically applied to reduce the spatial fluctuations further and to form a dipolar chain through the strong dipolar interaction. Finally, the in-plane optical lattice is adiabatically reduced to zero, leading to a chain liquid (or crystal) in the free 2D space. Evasion from the large fluctuation regime (i.e. weak lattice and weak field, see figure \ref{fig1}(d)) ensures that a strong collision rate between particles of different layers can form a chain.

To investigate the many-body properties, we consider the regime where these chains form a crystal phase with a small spatial fluctuation. By explicitly calculating the phonon spectrum and the quantum zero point energy, we can obtain the system entropy and monitor how the system temperature changes during the formation of chains. Furthermore, we calculate the system free energy and observe a quantum dissociation phase transition from a classical chain crystal to layers of crystal, i.e. the binding particles within a chain becomes unbound, as inter-layer distance is larger than a critical value. This dissociation transition is driven by the many-body quantum fluctuation effects (i.e. not existing in a single dipolar chain), and leads to a divergent compressibility along the chain at the phase boundary. Such geometrically induced quantum phase transition is the feature of dipolar chain liquid/crystal, resulting from the interplay of anisotropic dipolar interaction with anisotropic trapping potential. We further discuss how such quantum phase transition can be observed in the present experimental parameter regime.

The paper is organized as following: In section \ref{hamiltonian} we first present the system Hamiltonian and its harmonic approximation within the small fluctuation limits. In section \ref{phonon}, we show the phonon excitation spectrum for various of parameter regimes. In section \ref{temperature} we present how the temperature changes during the adiabatic (constant entropy) process as shown in figure \ref{fig1}(d). In section \ref{many-body}, we further calculate some many-body properties of the system, including entropy, free energy, system pressure, and compressibility from these elementary excitations. Using these results, we demonstrate the quantum phase transition from a chain crystal to layers of crystal. We then discuss the experimental relevant issues in section \ref{experiment} and summarize our results in section \ref{summary}.

\section{System Hamiltonian and harmonic approximation}
\label{hamiltonian}

A complete theoretical study on the formation and on the many-body properties of composite particles, like dipolar chains, is very challenging. To simplify the calculation, in this paper, we will limit our calculation to the regime of small spatial fluctuations (i.e. spatial variation is smaller than average inter-particle distance), resulting from either a strong lattice potential and/or strong dipole interaction (see figure \ref{fig1}(d)). In other words, throughout this paper, we always assume the spatial fluctuation is so small that dipoles are always in a crystal phase rather than in a liquid state, although most of our results should also applied to the liquid phase at least qualitatively.

We first consider dipolar atoms/molecules loaded in a stack of 2D pancake layers, where additional in-plane triangular optical lattice is applied (see figure \ref{fig1}(a)). From theoretical point of view, such initial preparation can simplify the calculation greatly since the average positions of each particles will not be changed during the whole process of manipulating the external field and in-plane lattice potential (the particle density is assumed fixed). Extension to other types of in-plane lattice potential is in principle possible, but the theoretical calculation becomes much more complicated, because the equilibrium position of dipolar particles can be changed as the in-plane lattice is reduced, in order to form a self-assembled dipolar crystal. Since the thermodynamic properties is independent of the initial parameter during the adiabatic process, we believe the details of the initial preparation should not affect the final conclusion of our results.

We start from a system, where polar molecules (or other dipolar particles) are initial prepared in a Mott insulator state with one particle per site, as described in figure \ref{fig1}(a), the system Hamiltonian including the the dipolar interaction can be written as
\be
H&=&\sum_{\bfn} \frac{\bfp_{\bfn}^2}{2m}
+\frac{1}{2}\sum_{n_3=m_3}\sum_{(n_1,n_2)\neq (m_1,m_2)}V_{\|}({\bf R}_{\bfn}-{\bf R}_{\bfm})\nonumber\\
&+&\frac{1}{2}\sum_{n_{3}\neq m_{3}}\sum_{(n_1,n_2)\neq (m_1,m_2)}V_{\perp}({\bf R}_{\bfn}-{\bf R}_{\bfm})
+\sum_{\bfn} V_{\rm lat}({\bf R}_{\bfn})\nonumber\\
\ee
where ${\bf R}_{n}=n_1a\hat{\bfa}_1+n_2a\hat{\bfa}_2+n_3 d\hat{z} +\bfr_{\bfn}$ is the position of particles at the lattice coordinate, $\bfn\equiv (n_1,n_2,n_3)$. $\hat{\bfa}_1=(1,0)$ and $\hat{\bfa}_2=(1/2,\sqrt{3}/2)$ are the two unit vectors for the in-plane triangular lattice and $\hat{z}$ is the unit vector perpendicular to the plane. $a$ and $d$ are their lattice constants respectively, leading to the two energy unit defined as $ E_a\equiv \hbar^2/m a^2$ or $ E_0\equiv\hbar^2/m d^2=\frac{1}{2\pi^2}E_R$, where $E_R$ is the recoil energy. $\bfr_{\bfn}$ is the in-plane spatial variation from the lattice center, and  $V_{\|,\perp}({\bf R}_n-{\bf R}_m)$ are the intra-plane and inter-plane interaction respectively. Finally, $V_{\rm lat}$ is the in-plane lattice potential. For simplicity, here we have neglected the fluctuation in the perpendicular direction ($\hat{z}$) by assuming a very strong 1D optical lattice.

Since we are interested in the regime of small spatial fluctuations, we can expend the above Hamiltonian to the quadratic order about the equilibrium lattice positions, leading to $H=H_{ph}+E_{\|}+E_{\perp}$, where the last two terms are the in-plane and out-of-plane classic energies (see Appendix \ref{appendix} for explicit expressions), and the quadratic Hamiltonian ($H_{ph}$) can be easily diagonalized. The phonon excitation can be expressed to be
%------
\begin{equation}
\hbar \Omega_{\alpha}(\bfk)= \sqrt{(\hbar\omega)^2+U_{dd}E_0\lambda_{\bfk,\alpha}}
\end{equation}
%------
where $U_{dd}\equiv D^2/d^3$ measures the strength of dipolar interaction with the electric dipole moment, $D$, and $\lambda_{\bfk,\alpha}$ is the eigenvalue of the interaction matrix element of the quadratic order (see Appendix \ref{appendix} for explicit expressions).
$\omega$ is the effective harmonic frequency inside each optical lattice, obtained by expanding $V_{\rm lat}({\bf R}_{\bfn})$ with respect to the lattice point to the quadratic order.

\section{Phonon spectrum of dipolar chain crystal in 2D free space}
\label{phonon}
%------
\begin{figure}[htbp!]
\centering
\includegraphics[width=8.5cm]{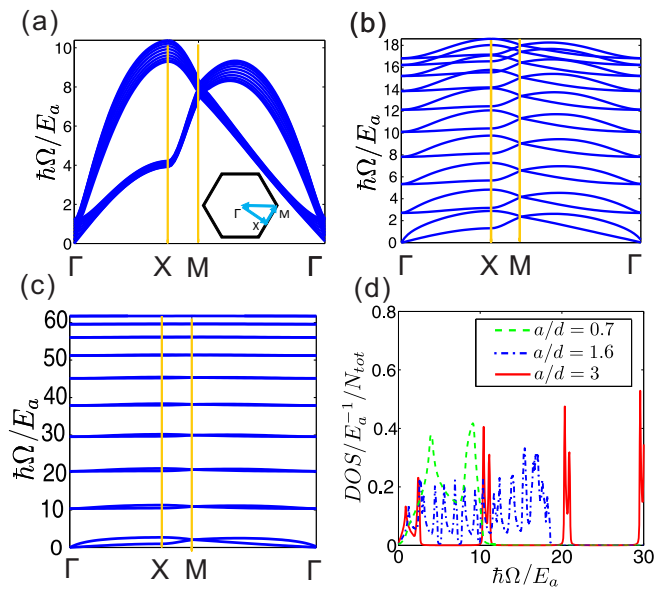}
\caption{(a), (b) and (c) show the typical phonon dispersion of self-assembled dipolar chain crystal along high symmetry points in momentum space for different in-plane density ($a/d=0.7,$ $a/d=1.6$ and $a/d=3$ respective). Here we consider $M= 10$ layers in a uniform space ($\hbar\omega=0$) with dipolar interaction $U_{dd}/E_0=1$. (d) shows the density of states for the three cases.}
\label{fig2}
\end{figure}
%---------

In figure \ref{fig2}, we show the calculated in-plane phonon excitation dispersion for a system of dipolar chain crystal in free space (i.e. no in-plane lattice potential, $\omega=0$). We consider ten layers ($M=10$) with three different lattice constants (i.e. three different densities): $a/d=0.7$, 1.6, and 3. The high symmetry points $\Gamma$, $X$ and $M$ in the reciprocal lattice are depicted in the inset of figure \ref{fig2}(a). Note that smaller $a/d$ leads to the single layer limit with ten-fold degeneracy in each 2D plane mode. In this limit, the properties of a chain is poorly-defined, because any in-plane excitation of finite momentum can be larger than the internal excitation energy of a chain, leading to a strong in-plane correlation. On the other hand, larger $a/d$ in a dilute limit is close to a single chain limit, where the discrete energy band in figure \ref{fig2}(c) shows the eigen-modes for a quantized vibration of chains. Note that the in-plane coupling does broaden the eigenmodes of the chain vibration to form a band, while a single chain is still well-defined since the band width is smaller than the energy gap in a chain, i.e. when applying an in-plane local measurement to identify a single chain, the energy perturbation can be (in principle) so small that no internal degree of chains is excited.

In figure \ref{fig2}(d), we further show structure of the density of state (DOS) for various $a/d$. It is easy to see that when $a/d$ is smaller (higher density limit) there are mainly two acoustic phonon bands, dominated by the excitation energy near $X$ point. These two energy bands indicate two temperature scales when thermal excitations are occupied from the low temperature limit. This may be related to the two-stage melting process (solid-hexatic-liquid) for a 2D triangular lattice, which has been extensively discussed in the literature\cite{Kosterlitz,Halperin,Nelson,Young,Strandburg}.

When $a/d$ is larger, the excitations along the chain direction become comparable to the in-plane ones, lifting the degeneracy in the DOS by  showing more bands for each mode along the chain. When $a/d\sim 1.6$, the DOS starts to have gaps opened at finite energy. The presence of narrow bands with gaps between them shows the possibility to identify a single mode excitation along the chain direction in such strongly interacting chain crystal. Therefore, when a weakly coupled 2D crystal (small $a/d$) is adiabatically changed to a weakly coupled chain crystal (large $a/d$), we do expect a significant change in the system properties (see below).

From the phonon excitations, we can then calculate the spatial fluctuations of each dipolar particles due to quantum zero point energy and/or thermal energy. They have to be smaller than the inter-particle distance during the whole process (see figure \ref{fig1}(d)) in order to justify the calculation of phonon excitations. Our calculation shows that this criteria is well-fulfilled in all the parameter range we considered here and therefore will not mention this again throughout this paper.m.

\section{Entropy and Temperature change during the chain formation}
\label{temperature}
%-----------
\begin{figure}[htbp!]
\centering
\includegraphics[width=9cm]{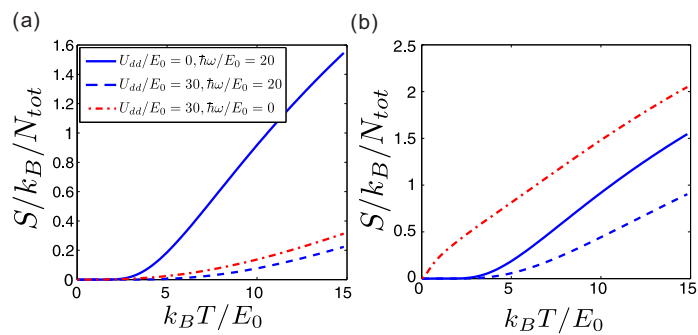}
\caption{The entropy per particle as a function of temperature for a system of 10 layers with a lattice constant (a) $a/d=0.7$, and (b)$a/d=3$. Solid lines are results near the stage 1 of figure \ref{fig1} with a strong optical lattice potential but zero dipole moment (no external field). Dash lines are results of the stage 2 with both a strong lattice potential and a strong dipole moment. Red dash-dotted lines are results of the stage 3 with a strong dipole moment but no in-plane lattice potential (free space).}
\label{fig3}
\end{figure}
%----------
%-----------
\begin{figure}[htbp!]
\centering
\includegraphics[width=9cm]{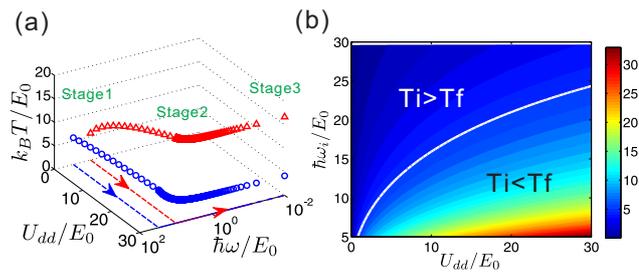}
\caption{ (a) Change of the system temperature during the adiabatic process (stage 1 to stage 3 of figure \ref{fig1}) with lattice constant $a/d=0.6$, initial temperature $T_i=5.7E_0$, and layer number $M=10$. Different initial lattice strength can lead to cooling (blue circle, $\hbar\omega/E_0=30$) or heating (red triangle, $\hbar\omega/E_0=10$)
in the final stage. (b) shows the density plot of the final temperature (for the same initial temperature) as a function of the lattice strength and dipolar interaction strength of the stage 2 in figure \ref{fig1}(d). The solid white line indicates the parameters to have final state temperature the same as the initial temperature.}
\label{fig4}
\end{figure}
%----------

The change of phonon spectrum (see figure \ref{fig2}) leads to a significant change of thermodynamic properties. One of the most important questions is how the system temperature changes during the whole process of figure \ref{fig1}, if it is kept adiabatic (constant entropy). The system entropy can be easily calculated from the phonon excitation spectrum through the partition function, ${\cal Z}\equiv \textmd{Tr} e^{- H/k_B T}$, so that
$S/k_B=\ln {\cal Z}+E_{\rm tot}/k_BT$, where the quantum zero point energy of the phonon fluctuations has been included in the total energy, $E_{\rm tot}$.

In figure \ref{fig3}, we show how the entropy changes as a function of temperature for various parameter. There are two general features: First, for a fix lattice potential, systems of a stronger dipole interaction always have lower entropies (i.e. entropy decreases from stage 1 to stage 2). This can be easily understood, since the dipole interaction will frozen the relative motion between dipoles to form a crystal phase with less spatial variation. Secondly, for a given dipolar interaction, systems of a weak in-plane lattice potential always have larger entropy (i.e. entropy increases from stage 2 to stage 3), since the spatial variation increases when the optical lattice potential decreases. As a results, the final temperature can be either increased (heating) or decreased (cooling), determined by the competition of these two processes.

However, as mentioned above, the phonon spectrum of a free standing dipolar chain (i.e. without in-plane optical lattice) strongly depends on the in-plane density of chains. For systems of higher density (smaller $a/d$, see dash-dotted lines of figure \ref{fig3}(a)), the entropy increases as $T^2$ in the low temperature limit, contributed from the linear dispersion of the in-plane phonon excitations as shown in a single layer crystal limit (figure \ref{fig2}(a)). In the dilute limit with a discrete energy band (see figure \ref{fig2}(c)), on the other hand, the low temperature behaviour changes more dramatically (see figure \ref{fig3}(b)). Such different low temperature behaviour of entropy leads to a significant results of the temperature change during the formation of chain crystal: If dipoles are initially prepared in a dilute limit, the entropy of dipolar chain crystal without optical lattice (stage 3, dash-dotted lines) is larger than the entropy of the initial state for particles confined in the optical lattice without (or with weak) dipole moment (stage 1, solid lines), making the whole process to be a cooling process. This effects, however, can be in the opposite (heating) if the in-plane density of chains is larger as shown in figure \ref{fig3}(a).

As for the adiabatic process by conserving the system entropy, in figure \ref{fig4}(a) we show how the system temperature changes during the three stages to form dipolar chain crystal. When DC field is gradually increased from $U_{dd}/E_0=0$ (stage 1) to $U_{dd}/E_0=30$ (stage 2), the level spacing is enhanced and the raise of temperature is followed to keep the same number of energy levels. The temperature then decreases when the in-plane lattice potential is reduced to zero as shown above. In figure \ref{fig4}(b), we show how the final temperature depends on the lattice potential and/or dipolar interaction strength of the intermediate stage (stage 2). If the initial lattice potential is strong enough in stage 1, the final state temperature can be lowered than the initial temperature, leading to a polarization cooling as also observed in a dipolar system confined in a harmonic potential \cite{Tian}.

\section{Quantum Dissociation Phase Transition of a Chain Crystal}
\label{many-body}

It is known that the ground state of a multi-layer system of one particle per layer should be a bound state because of the attractive interaction between dipolar atoms/molecules\cite{Shih,Armstrong2,Klawunn2}, while mutual repulsive interaction between dipoles in the same layer certainly increases the energy cost and may lead to a dissociation transition. Similar situations also appear in other systems of composite particles, for example, transition between Wannier exciton (bound state) and electron-hole plasma (unbound state) \cite{Zimmermann,Forchel,Zimmermann1,DasSarma}, crossover physics between BEC (bound state) and BCS (Cooper pair) for fermions in free space\cite{Leggett,Nozieres,Engelbrecht,Bloch2,Giorgini} or in bilayer systems\cite{Pikovski,Zinner} etc.

To investigate the many-body problem between the bound and unbound states, or the possible dissociation transition of a dipolar chain crystal, we first calculate the Helmholtz free energy, $F(T,N)=-k_BT\ln{\cal Z}(N,T)$, and then the "pressure" along $z$ direction, $P_{z}=-\partial F/\partial V$ (here $V=A M d$ is the volume with $A$ and $M$ being the layer area and the number of layers respectively. Both $A$ and $M$ are fixed for the differential). The isothermal compressibility along the chain direction can be also defined to be, $\beta^z_T\equiv -V^{-1}\partial V/\partial P_z$, which reflects the response to the variation of the inter-layer spacing along the chain direction.

%-----------
\begin{figure}[htbp!]
\centering
\includegraphics[width=9cm]{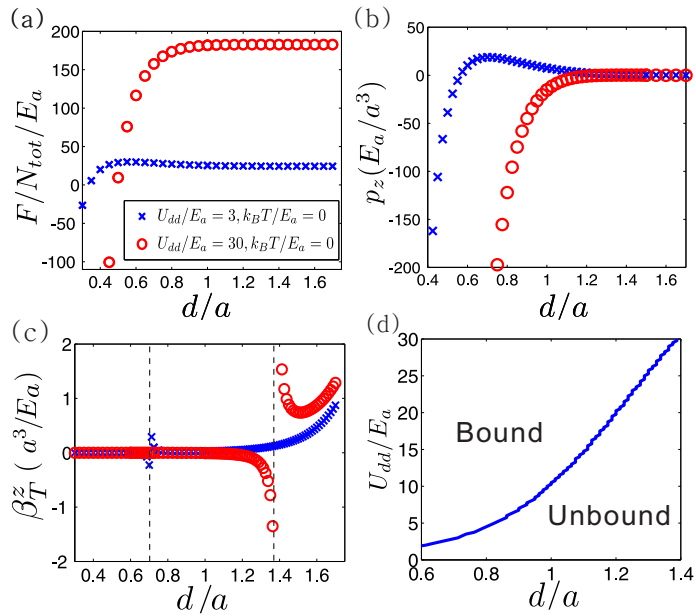}
\caption{(a) Helmholtz free energy ($F$) per particle, (b) System pressure ($P_z$) along the normal direction, and (c) the isothermal compressibility ($\beta_T$) as a function of inter-layer distance, $d/a$. Results of two different dipole interaction energies are shown together for comparison. The two vertical thin dashed lines to indicate the position of phase boundary, defined by $d_c/a$, where $\beta^z_T$ diverges.(d) Phase diagram of a dipolar chain crystal as a function of inter-layer distance ($d/a$) and dipolar interaction ($U_{dd}/E_a$). All the results are calculated for a system of ten layers at zero temperature. Results of more layers are almost the same.
}
\label{fig5}
\end{figure}
%----------

In figure \ref{fig5}(a), we show the Helmholtz free energy calculated at zero temperature (i.e. equivalent to the total energy). It is easy to see that in short inter-layer distance limit, the free energy is dominated by the classical energy (i.e. $E_{\perp}+E_{\|}$), which is purely attractive and scaled as $-(1/d)^{3}$. When the inter-layer distance is larger, however, quantum zero point energy due to phonon fluctuations  becomes dominant, because it scales as $const+d^{-5/2}$ in the large $d/a$ limit (the constant term results from the in-plan quantum fluctuation). The competition between the classical and the quantum zero point energies makes such anisotropic systems highly non-trivial: the pressure along $z$ direction reaches its maximum at $d=d_c\sim a$ and becomes negative at shorter distances. The maximum of pressure indicates the divergence of compressibility, $\beta^z_T$, at $d=d_c$. This quantum dissociation phase transition is second order with a typical mean-field scaling exponent.

When the dipolar chain crystal becomes dissociated due to quantum fluctuation, the ground state of the system can be regarded as a collection of a 2D dipolar crystal in each layer, self-assembled by the repulsive intra-layer interaction as extensively studied before. The strong phonon fluctuations within each layer makes the interlayer coherence to be short-ranged, very different from a regular dipolar chain liquid/crystal.
Note that the dissociation phase transition above is a pure "many-body" effect due to the interaction between particles in the chain crystal. Such transition does not exist if considering a single chain only, which is shown always the ground state at zero temperature as have been studied before\cite{DW, Zhu, Klawunn}. Besides, even near the phase transition boundary, the spatial fluctuation of all dipolar particles about its average position can be still small so that the calculation of phonon spectrum in this paper is still well-justified. In figure \ref{fig5}(d), we further show the quantum phase diagram as a function of inter-layer distance and the dipole interaction strength. Results for more than ten layers are very similar, since the many-body properties are mainly determined by dipoles in the bulk. Finally, due to the anisotropic nature of dipolar interaction, the in-plane pressure and the in-plane compressibility can be very different from the results perpendicular to the layer (i.e. along the chain direction). According to our calculation, in-plane pressure(compressibility) is monotonically decreasing(increasing) as a function of inter-chain distance, $a/d$, and therefore we will not show it here.

\section{Experimental parameters and measurement}
\label{experiment}

Before conclusion, we discuss the experimentally relevant issues here. We first consider polar molecules like LiCs as an example. For LiCs \cite{Deiglmayr}, the typical energy scale, $E_0= \hbar^2/m d^2\sim 257$ Hz for an optical lattice of lattice spacing $\lambda/2=530$ nm(Recoil energy $E_R \sim 1.27$ kHz). The corresponding dimensionless strength of the dipole interaction $U_{dd}/E_0=mD^2 /\hbar^2d$ can be as much as $15.8$, when the electric dipole moment $D=2$ Debye (the permanent dipole
moment of this molecule in the ground state is about $D = 5.5
$ Debye). As a result, we can easily estimate the critical temperature of a single layer dipolar crystal to be \cite{Kalia,Buchler}: $T_c\sim 0.09  D^2/((3/4)^{1/4} a)^3\sim 12 $ nK for $a\sim d$. This critical temperature should be achievable in our scheme if the initial temperature inside the optical lattice to be about $1.75 E_0\sim 21$ nK with a lattice strength $V_0 \sim 3$ kHz (where $\hbar\omega=2 \sqrt{V_0 E_r}=15 E_0$), or to be about $3.5 E_0\sim 42$ nK with a lattice strength $V_0\sim 12$ kHz (where $\hbar\omega=30 E_0$).

For magnetic atoms like Dy \cite{Benjamin}, the energy scale $E_0\sim221$ Hz, but the interaction energy, $U_{dd}/E_0=0.04$, is very small even for a fully polarized magnetic moment, $\mu=10\mu_B$. As a result, we do not expect a promising results of chaining or crystallization for magnetic dipolar atoms, although certain long-ranged interaction effects should be still observable.

For the measurement of dipolar chains, spectroscopic techniques, such as lattice shaking\cite{Thilo, Schori}, should be available for measuring the bound state energy. Besides, since the effective mass of a chain is much larger than the original molecular mass, {\it in situ} measurement and/or the change of dynamical properties can be a direct evidence of the composite particles (i.e. chaining). Time-of-flight measurement should provide a direct evidence of the crystal order, and the interference should be averaged out when the chains are dissociated and the inter-layer correlation disappear. Rf spectroscopy \cite{Stewart} and Bragg spectroscopy\cite{Stenger, StamperKurn} etc. could provide additional information on the phonon dispersion and density of states of the chain crystal.

In order to observe the quantum dissociation phase transition discussed in this paper, we propose that one can also measure the response of the system with respect to the perturbation of confinement potential along the chain direction (perpendicular to the layer). Since the isothermal compressibility along the chain direction should diverges near the phase boundary, it is nature to expect a highly non-trivial quadrupole collective motion along the chain direction, if the inter-layer single particle tunnelling amplitude can be tuned from zero to a finite but small value (making the dynamics along the chain direction available). Furthermore, according to the fluctuation-dissipation theorem, the isothermal compressibility is proportional to the number fluctuations within a grand canonical ensemble \cite{Pathria}, i.e.
\be
\langle \Delta N_{tot}^2\rangle=\langle N_{tot}^2\rangle-\langle N_{tot}\rangle^2=\frac{\langle N_{tot}\rangle ^2}{V}\times k_BT\kappa_T,
\ee
we may expect a large number fluctuation in each layer when the parameters are tuned near the phase boundary ($\kappa_T\to\pm\infty$). This can be easily observed from {\it in situ} measurement or interference pattern when a small but finite inter-layer tunneling is turned on.

Finally, we emphasize that such quantum dissociation should exist even in a dipolar chain liquid phase, because strong intra-layer quantum fluctuations are more important. Finite temperature calculation should also reveal the nontrivial change of the critical temperature when the chains of crystal melt. These liquid phases (either normal or quantum) cannot be correctly described in the harmonic expansion of this paper, because the spatial fluctuations can become the same order as the inter particle distance. Full quantum mechanical simulation (real space quantum Monte Carlo, for example) will be needed to investigate the full phase diagram.

%%%%%%%%%%%%
\section{Summary}
\label{summary}

In summary, we proposed an experimental scheme to effectively assemble dipolar gases (especially polar molecules) to form dipolar chains, and calculate the evolution of temperature of whole adiabatic manipulation. We also calculate the collective modes (phonon excitations) of a dipolar crystal, and demonstrate a quantum dissociation phase transition between two distinct phases: dipolar chain crystal and layers of dipolar crystal. The quantum zero point energy leads to a divergence of the compressibility near the phase boundary. The formation of dipolar chains, phonon excitations of chain crystal and the dissociation phase transition should be easily observable within present experimental techniques, opening a new area for strongly interacting dipolar gases.

%%%%%%%%%%%%
\section{Acknowledgement}

We thank fruitful discussion with C.-H. Chen, E. Demler, K.-K. Ni, Y.-Y. Tian, and J. Ye. This work was supported by National Center for Theoretical Sciences and National Sciences Council in Taiwan.  JS You also acknowledges the support from NSC Grant No. 102-2917-I-007-032.

%%%%%%%%%%%%%%%%%%%%%%%%%%%%%%
%%%%%%%%%%%%%%%%%%%
\appendix
\section{Hamiltonian}
\label{appendix}
The Hamiltonian including the dipolar interaction and optical lattice can be described by
\begin{small}
\be
\fl{
H=\sum_{\bfn} \frac{\bfp_{\bfn}^2}{2m}
+\frac{1}{2}\sum_{n_3=m_3}\sum_{n_{\rho}\neq m_{\rho}}V_{\|}({\bf R}_{n}-{\bf R}_{m})
+\frac{1}{2}\sum_{n_{3}\neq m_{3}}\sum_{n_{\rho},m_{\rho}}V_{\perp}({\bf R}_{n}-{\bf R}_{m})+\sum_{\bfn} V_{lat}({\bf R}_{n}),}\nonumber\\
\ee
\end{small}
%----
where the in-plane interaction matrix element is
\begin{small}
\begin{flalign}
V_\|({\bf R}_{n_{\rho},n_3}^{(0)}-{\bf R}_{m_{\rho},n_3}^{(0)}+\bfr_{n_{\rho},n_3}-\bfr_{m_{\rho},n_3})&=
\frac{D^2/d^3}{\left|(n_1-m_1)\bfa_1+(n_2-m_2)\bfa_2+\Delta \bfr^\|_{\bfn\bfm}
\right|^3}
\nonumber\\
&=\frac{D^2/d^3}{l_{\bfn,\bfm}^3}
\times\left[1-\frac{3}{2l_{\bfn,\bfm}^2}\Delta\bfr^\|_{\bfn,\bfm}{}^2
+\frac{15}{2}
\frac{1}{l_{\bfn,\bfm}^4}\left({\bf l}_{\bfn,\bfm}\cdot
\Delta\bfr_{\bfn,\bfm}^\|\right)^2+\cdots\right]
\end{flalign}
\end{small}
%----
and the inter-layer interaction matrix element is
%-----
\begin{small}
\begin{flalign}
&V_\perp({\bf R}_{\bfn}^{(0)}-{\bf R}_{\bfm}^{(0)}
+\bfr_{\bfn}-\bfr_{\bfm})=
\frac{D^2/d^3}{\left|\bfl_{\bfn,\bfm}+\Delta \bfr^\perp_{\bfn\bfm}
+h_{\bfn,\bfm}\hat{z}\right|^3}
\left(1-\frac{3 h_{\bfn,\bfm}^2}{|\bfl_{\bfn,\bfm}+\Delta\bfr^\perp_{\bfn,\bfm}
+h_{\bfn,\bfm}\hat{z}|^2}\right)
\nonumber\\
&=\frac{D^2}{d^3}\left[\frac{-2h_{\bfn,\bfm}^2+l_{\bfn,\bfm}^2}{(l_{\bfn,\bfm}^2+h_{\bfn,\bfm}^2)^{5/2}}
+\frac{3(4h_{\bfn,\bfm}^2-l_{\bfn,\bfm}^2)}{2(l_{\bfn,\bfm}^2+h_{\bfn,\bfm}^2)^{7/2}}
\Delta\bfr^\perp_{\bfn,\bfm}{}^2+
\frac{15(-6h_{\bfn,\bfm}^2+l_{\bfn,\bfm}^2)}{2(l_{\bfn,\bfm}^2+h_{\bfn,\bfm}^2)^2}
\left({\bf l}_{\bfn,\bfm}\cdot
\Delta\bfr^\perp_{\bfn,\bfm}\right)^2+\cdots\right]
\end{flalign}
\end{small}
Here we have defined the relative coordinate (normalized to the inter-layer distance, $d$): ${\bf l}_{\bfn,\bfm}=({\bf R}_{n_{\rho},n_3}^{(0)}-{\bf R}_{m_{\rho},n_3}^{(0)})/d=((n_1-m_1)\bfa_1+(n_2-m_2)\bfa_2)/d$ with $\bfa_1=a(1,0), \bfa_2=a(1/2, \sqrt{3}/2)$ , ${h}_{\bfn,\bfm}=(n_3-m_3)$, $\Delta\bfr_{\bfn,\bfm}^\|\equiv (\bfr_{n_{\rho},n_3}-\bfr_{m_{\rho},m_3=n_3})/d$ and $\Delta\bfr_{\bfn,\bfm}^\perp\equiv (\bfr_{\bfn}-\bfr_{\bfm})/d$ (for $n_3\neq m_3$). We focus on deviations of molecules from their equilibrium positions, so that the terms linearly proportional to  $\Delta\bfr_{\bfn,\bfm}^{\|,\perp}$ must be summed to zero.

By keeping the Hamiltonian to quadratic order of $r_{\bfn}$  and using the Fourier transformation in $x-y$ plane  $\bfr_\bfn/d=\frac{1}{\sqrt{N_{2D}}}\sum_{\bfk}\bfu_{\bfk}(n_3)\,e^{i\bfk\cdot\bfl_\bfn d}$ with $N_{tot}=N_{2D} M$, we have
\be
H&=&H_{ph}+E_{\|}+E_{\perp},
\ee
where
\be
E_{\|}&=&\frac{D^2}{2d^3}\sum_{n_{3}, m_{3}}\sum_{n_{\rho}\neq m_{\rho}} \frac{1}{l_{\bfn,\bfm}^3},\\
E_{\perp}&=&\frac{D^2}{2d^3}\sum_{n_{3}\neq m_{3}}\sum_{n_{\rho},m_{\rho}}\frac{-2h_{\bfn,\bfm}^2+l_{\bfn,\bfm}^2}{(l_{\bfn,\bfm}^2+h_{\bfn,\bfm}^2)^{5/2}}
\ee
and
\be
H_{ph}&=&\frac{1}{2m}\sum_{\bfn} \bfp_{\bfn}^2+\frac{m\omega^2}{2}\sum_{\bfn} {\bf r}_{\bfn}^2\nonumber+\frac{D^2}{2d^3} \sum_{\bfk}\bfu_\bfk^\dagger V_{\bfk} \bfu_\bfk\nonumber\\
\ee
with the following dimensionless interaction matrix in the basis of lattice fluctuations:

\be
 V_{\bfk}=\begin{bmatrix}
V^{xx}_\bfk & V^{xy}_\bfk   \\
V^{yx}_\bfk & V^{yy}_\bfk
\end{bmatrix},
\ee
and
\be
\bfu_\bfk^\dagger
=\begin{bmatrix}
   u^*_{x,\bfk}(1),\cdots,u^*_{x,\bfk}(M), & u^*_{y,\bfk}(1),\cdots,u^*_{y,\bfk}(M)
   \end{bmatrix}.
   \ee
The diagonal elements of $V^{xx}_\bfk, V^{xy}_\bfk$ and $V^{yy}_\bfk$ are shown below:
\begin{small}
\begin{flalign}
V^{xx}_\bfk(n_3,n_3)&=\sum_{\bfp\neq 0}\frac{\sin^2( \bfk\cdot\bfl_\bfp/2)}{|\bfl_\bfp|^5}
\left(-6+\frac{30(p_1a_1^x+p_2a_1^x)^2}{|\bfl_\bfp|^2}\right)\nonumber\\
&+\sum_{m_3,m_3\neq n_3}\sum_{\bfp}\left[\frac{12{h}_{\bfn,\bfm}^2-3|\bfl_\bfp|^2}{({h}_{\bfn,\bfm}^2+|\bfl_\bfp|^2)^{7/2}}-\frac{90{h}_{\bfn,\bfm}^2-15|\bfl_\bfp|^2}{({h}_{\bfn,\bfm}^2+|\bfl_\bfp|^2)^{9/2}}(p_1a_1^x+p_2a_2^x)^2\right],
\\
V^{yy}_\bfk(n_3,n_3)&=\sum_{\bfp\neq 0}\frac{\sin^2( \bfk\cdot\bfl_\bfp/2)}{|\bfl_\bfp|^5}
\left(-6+\frac{30(p_1a_1^y+p_2a_1^y)^2}{|\bfl_\bfp|^2}\right)\nonumber\\
&+\sum_{m_3,m_3\neq n_3}\sum_{\bfp}\left[\frac{12{h}_{\bfn,\bfm}^2-3|\bfl_\bfp|^2}{({h}_{\bfn,\bfm}^2+|\bfl_\bfp|^2)^{7/2}}-\frac{90{h}_{\bfn,\bfm}^2-15|\bfl_\bfp|^2}{({h}_{\bfn,\bfm}^2+|\bfl_\bfp|^2)^{9/2}}(p_1a_1^y+p_2a_2^y)^2\right],
\\
V^{xy}_\bfk(n_3,n_3)&=\sum_{\bfp\neq 0}
30(p_1a_1^x+p_2a_1^x)(p_1a_1^y+p_2a_1^y)\frac{ \sin^2( \bfk\cdot\bfl_\bfp/2)}{|\bfl_\bfp|^7}\nonumber\\
&+\sum_{m_3,m_3\neq n_3}\sum_{\bfp}\left[\frac{-90{h}_{\bfn,\bfm}^2+15|\bfl_\bfp|^2}{({h}_{\bfn,\bfm}^2+|\bfl_\bfp|^2)^{9/2}}(p_1a_1^x+p_2a_2^x)(p_1a_1^y+p_2a_2^y)\right]=V_\bfk^{yx}(n_3,n_3).
\end{flalign}
\end{small}

For the off-diagonal elements of $V^{xx}_\bfk, V^{xy}_\bfk$ and $V^{yy}_\bfk$, we have

\begin{small}
\begin{flalign}
V^{xx}_\bfk(n_3,m_3)&=-\sum_{\bfp}\left[\frac{12{h}_{\bfn,\bfm}^2-3|\bfl_\bfp|^2}{({h}_{\bfn,\bfm}^2+|\bfl_\bfp|^2)^{7/2}}-\frac{90{h}_{\bfn,\bfm}^2-15|\bfl_\bfp|^2}{({h}_{\bfn,\bfm}^2+|\bfl_\bfp|^2)^{9/2}}(p_1a_1^x+p_2a_2^x)^2\right]\cos ( \bfk\cdot\bfl_\bfp),
\\
V^{yy}_\bfk(n_3,m_3)&=-\sum_{\bfp}\left[\frac{12{h}_{\bfn,\bfm}^2-3|\bfl_\bfp|^2}{({h}_{\bfn,\bfm}^2+|\bfl_\bfp|^2)^{7/2}}-\frac{90{h}_{\bfn,\bfm}^2-15|\bfl_\bfp|^2}{({h}_{\bfn,\bfm}^2+|\bfl_\bfp|^2)^{9/2}}(p_1a_1^y+p_2a_2^y)^2\right]\cos ( \bfk\cdot\bfl_\bfp),
\\
V^{xy}_\bfk(n_3,m_3)&=V^{yx}(n_3,m_3)=\sum_{\bfp}\left[\frac{90{h}_{\bfn,\bfm}^2-15|\bfl_\bfp|^2}{({h}_{\bfn,\bfm}^2+|\bfl_\bfp|^2)^{9/2}}(p_1a_1^x+p_2a_2^x)(p_1a_1^y+p_2a_2^y)\right]\cos ( \bfk\cdot\bfl_\bfp).
\end{flalign}
\end{small}
%

%%%%%%%%%%%%%%%%%%%%%%%%%%%%%%

\end{document}